\documentclass{llncs}
\usepackage[pdftex]{graphicx}
\usepackage{epstopdf}        
     
\usepackage{makeidx}  
\usepackage{amsmath}
\usepackage{amsfonts,amssymb}
\usepackage{clrscode3e}        
\usepackage{multirow}

\begin{document}
\frontmatter          
\pagestyle{headings}  
\addtocmark{Connectivity of a graph} 
\title{Finding the connected components of the graph\\ using perturbations of the adjacency matrix}
\titlerunning{Connectivity of a graph}  
%
\author{Alexander Prolubnikov}
\authorrunning{Alexander Prolubnikov} 
%
\tocauthor{Alexander Prolubnikov}
\institute{Omsk State University, Omsk, Russian Federation\\
\email{a.v.prolubnikov@mail.ru}
}
\maketitle              

\begin{abstract}
The problem of finding the connected components of a graph is considered. The algorithms addressed to solve the problem are used to  solve such problems on graphs as problems of finding points of articulation, bridges, maximin bridge, etc. A natural approach to solving this problem is a breadth-first search, the implementations of which are presented in software libraries designed to maximize the use of the capabi\-lities of modern computer architectures. We present an approach using perturbations of adjacency matrix of a graph. We check wether the graph is connected or not by comparing the solutions of the two systems of linear algebraic equations (SLAE): the first SLAE with a perturbed adjacency matrix of the graph and the second SLAE with~unperturbed matrix. This approach makes it possible to use effective numerical implementations of SLAE solution methods to solve connectivity problems on graphs. Iterations of iterative numerical methods for solving such SLAE can be considered as carrying out a graph traversal. Generally speaking, the traversal is not equivalent to the traversal that is carried out with breadth-first search. An algorithm for finding the connected components of a graph using such a traversal is presented. For any instance of the problem, this algorithm has no greater computational complexity than breadth-first search, and for~most individual problems it has less complexity.\\

\end{abstract}


\section{Introduction}
A lot of problems from applications related to infrastructure reliability issues such as transport net\-works, data transmission networks, large integrated circuits, sensor networks, etc. may be considered as connectivity problems on graphs. They are verification of a graph connectivity, finding its connected components, articulation points, bridges, etc. Often, the graphs for which these problems need to be solved in applications turn out to be sparse, that is, the number of their edges has the same order as the number of vertices: $m\!=\!O(n)$, where $n$ is the cardinality of the set of vertices of the graph, $m$ is the cardinality of its set of edges.

A natural approach to solve the connectivity problems on graphs is the implementation of {\it Breadth-First Search} (hereinafter we abbreviate it as {\it BFS}) \cite{Kormen}. This approach was first applied by K. Zuse in 1945, but was not published until 1972 \cite{Zuse}. Later, this algorithm was rediscovered in 1959 by E. Moore as an algorithm for finding the shortest path in the maze \cite{Moore} and it was later discovered independently of Moore and Zuse in the context of the problem of wiring conductors on printed circuit boards by Lee \cite{Lee}. To date, the use of BFS is a common choice when solving connectivity problems on graphs.

\smallskip

Before we start iterations of the algorithm implementing BFS, we choose some {\it starting} vertex from which the algorithm begins its work. At each iteration of the algorithm, we determine the vertices which are adjacent to those that we have already determined as  reachable from the starting vertex. Thus, during the operation of the algorithm, we sequential\-ly determine the levels of the {\it reachability tree}. Each vertex of the same level is reachable from the starting vertex in the same number of transitions along the edges of the graph. In one iteration of such an algorithm, one level of the reachability tree is obtained.

The computational complexity of BFS is $O(n\!+\!m)$, because during its iterations it is necessary to process $O(n)$ vertices and check for $O(m)$ edges. The computational complexity of BFS for an individual problem is affected by the choice of the starting vertex. The reachability tree for the connected component can be constructed in at least $\ell$ iterations, where $\ell$ is the diameter of the connected component.

Due to the inevitable sequential nature of computing of such vertex traversal, the construc\-tion of efficient parallel computing schemes for the implementation of BFS is difficult. For this reason, there are no sublinear implementations of BFS, i.e. having computational complexity $O(n^{c})$, where $0\!<\!c\!<\!1$.

\smallskip

By {\it algebraic} BFS is meant the implementation of BFS by successive multiplications of the vector by the adjacency matrix of the graph. That is, in the course of algebraic BFS application, we recursively obtain a sequence of vectors $x^{(k)}\!\in\!\mathbb{R}^n$, where $x^{(k+1)}\!=\!Ax^{(k)}$, $A$ is the adjacency matrix of the graph. $x^{(0)}\!=\!e_i\!=\!(0,\ldots,1,\ldots,0)$ is the $i^{th}$ basis vector of the standard basis of $\mathbb{R}^n$, where the only nonzero value is $1$ and it is placed in the $i^{th}$ position. Except for the last iteration of algebraic BFS, at each $k^{th}$ iteration of this process, the current vector $x^{(k)}$ obtains new non-zero componets. The set of such components forms the {\it level of the reachability tree}. The computational complexity of such an algorithm is minimal if the computational complexity of multiplication matrix by vector is $O(m)$. 

Algebraic BFS is implemented in software libraries designed to make the most effective use of the features of computer architectures and optimize work with memory. Such low-level implementations in practice can be faster than implementations of theoretically optimal combinatorial BFS \cite{BeamerAsanovicPatterson,BuckerSohr,BulucMattsonMcMillanMoreiraYang,AzadBuluc,YangBulucOwens,YangWangOwens,Burkhardt}.  For example, implementations of algebraic BFS for sparse graphs with their theoretical computational complexity equal to $O(mn)$ can give acceleration relative to having a lower theoretical complexity of BFS implemented through graph traversal. This is achieved by parallelizing calculations when obtaining one level of the reachability tree \cite{Burkhardt}.

\smallskip

We present an approach to finding the connected components of a graph using perturba\-tions of its adjacency matrix. The computational complexity of algorithms implemented this approach is $O(mn)$ without using any optimizing procedures for matrix multiplication. We check wether the graph is connected or not by comparing the solutions of the two systems of linear algebraic equations (SLAE): the first SLAE with a perturbed adjacency matrix of the graph and the second SLAE with~unperturbed matrix. This approach makes it possible to use effective numerical implementations of SLAE solution methods for solving connectivity problems on graphs, including the implementa\-tions of numerical methods for solving SLAE with sparse matrices.

Iterations of iterative numerical methods for solving SLAE can be considered as carrying out of a graph traversal, which, generally speaking, may not be equivalent to the traversal carried out by BFS. We present an algorithm for finding the connected components of a graph using such a traversal. The computational complexity of the algorithm is $O(m\ell)$, where $\ell$ is the graph diameter. For any instance of the problem, this algorithm has no greater computational complexity than algebraic BFS, and, for most individual problems, its computational complexity is less.\\

\section{Finding the connected components of a graph\\ using perturbations of the graph matrix}

An {\it undirected graph} $G$ is given by its set of vertices $V(G)$ and the set of edges $E(G)$, i.e. $G\!=\!\langle V(G),E(G)\rangle$, where $V(G)\!=\!\{1,\ldots,n\}$, $E(G)\subseteq V\!\times\! V$, and $(i,j)\!\in\! E(G)$ if and~only if $(j,i)\!\in\! E(G)$. Vertices $i,j\!\in\!V(G)$ are {\it adjacent} if $(i,j)\!\in\! E(G)$. Adjacent vertices $i$ and $j$ are said to be {\it incident} to the edge $(i,j)$. Further, each undirected edge $(i,j)$ is considered by us as the two oriented edges: the oriented edge $(i,j)$ and the oriented edge $(j,i)$.

A graph $G$ is called {\it connected} if for any two vertices from $V(G)$ there is a {\it chain} of edges from $E(G)$ connecting them. That is, there is a sequence of edges $(i_j,i_{j+1})$, $j\!=\!\overline{1,l}$, such that $i_1\!=\!i$, $i_l\!=\!j$. If $i\!=\!j$, then such a chain is called a {\it cycle}. {\it Cycle length} is the number of oriented edges from $E(G)$ forming the cycle. An undirected edge of the form $(i,i)$, $i\!\in\!V(G)$, is called a {\it loop}. Such loop is a cycle of length equal to zero.

Next, we shall consider only {\it simple} chains, that is, chains that do not contain repeating vertices.

The {\it connected component} of a graph is called the maximum connected subgraph of this graph by inclusion. For brevity, we shall also refer to the set of vertices of the component as a connected component.\\

\subsection{Modification of the adjacency matrix. Perturbations of the graph matrix} 

Let $A_0(G)$ is the {\it adjacency matrix} of a graph $G$, i.e. it has elements $$(A_0(G))_{ij}=\left\{
\begin{array}{ll}
1,&\mbox{ if } (i,j)\in E(G),\\
0,&\mbox{ else}.\\
\end{array}
\right.$$ By modifying $A_0(G)$, we get the matrix $A(G)$ of the graph $G$, which we shall use next: 
\begin{equation}\label{1}  
A(G)=A_0(G)+d\mathcal{I},
\end{equation}
where $\mathcal{I}$ is a unit matrix, and the values of the diagonal elements of $A(G)$ are equal to $d$ such that
\begin{equation}\label{2} 
d>\!\max_{i\in V(G)}d_i,
\end{equation}
where $d_i$ is the {\it degree} of the vertex $i\!\in\!V(G)$, that is, the number of edges incident to it from $E(G)$. Next, we shall assume $d\!=\!\mu\cdot d_{\max}$, where $\mu\!>\!1$. Thus the symmetric matrix $A(G)$ is positively defined and~has strict diagonal predominance. 

We consider matrices of the form (\ref{1}) as adjacency matrices of graphs with wighted loops. All of them are equal to $d$.

By {\it perturbation} of the matrix $A$ we mean getting the matrix $A'$ from it by changing its diagonal element:
\begin{equation}\label{3} 
A'=A+\varepsilon E_i,
\end{equation}
where $\varepsilon$ is the magnitude of the perturbation, $\varepsilon\!>\!0$, and the matrix $E_i$ has a single nonzero element in the $i^{th}$ position on the diagonal: $(E_i)_{ii}=1$, $(E_i)_{jk}=0$, $j\neq i$, $k\neq i$. The vertex $i$ is a {\it perturbed} vertex.

The matrix $A'$ can be considered as a matrix of the form (\ref{1}) of a graph in which the weight of the loop that is incident to vertex $i$ changes by $\varepsilon$.\\

\subsection{Finding the connected components of a graph} 

In order to determine the vertices belonging to the same connected component as the perturbed vertex $i$, having obtained the matrix $A'$, we find the inverse to it and analyze the response to the produced perturbation by which we mean the change in the values of elements of the inverse matrix as a result of the perturbation. Based on the presence or absence of this response in the elements of the $i^{th}$ column (or the $i^{th}$ row) of the inverse matrix, we find vertices belonging to the same connected component as the perturbed vertex. 

First, we find the solution $x$ of the SLAE
\begin{equation}\label{4} 
Ax=e_i.
\end{equation}
The solution $x$ is the $i^{th}$ column of the matrix $A^{-1}$. 

Then to find the vertices belonging to the same connected component as the perturbed vertex $i$, we perform the following steps.

\smallskip

\begin{itemize}
\item[1).] Solve the SLAE $Ax\!=\!e_i$.
\item[2).] Produce a perturbation of the vertex $i$: $A'\!=\!A+\varepsilon E_i$.
\item[3).] Solve the SLAE $A'x'\!=\!e_i$.
\item[4).] Compare the components of the obtained vectors $x'$ and $x$:\\ if $x_j'\!\neq\! x_j$, then the vertices $i$ and $j$ belong to the same connected component,\\ otherwise they belong to different connected component.
\end{itemize}

The notations $x$ and $x'$ for solutions of the SLAEs we shall use further in the text.

\smallskip

The algorithm $\proc{\bf ConnectedComponent\underline{s}}$ finds all of the connected components of an undirected graph according to this approach. It uses the procedure $\proc{\bf ConnectedComponent}$ to find one connected component to which the vertex $i$ belongs.

\begin{codebox}
\Procname{$\proc{\bf Connected component ($G, i\in\!V(G)):C$};$}\\
\li $C\leftarrow \{ i\}$; 
\li solve SLAE (\ref{2}), the solution is vector $x$;
\li solve SLAE (\ref{3}), the solution is vector $x'$;
\li \For $\forall j\in V$:
 \Do
\li \If $x_j'\neq x_j$
 \Then 
\li $C\leftarrow C\cup\{j\}$; 
\li $V\leftarrow V\setminus\{j\}$;
 \End
 \End\\
 
\li \mbox{\bf result}$\ \leftarrow C$. 
\end{codebox}

The algorithm $\proc{\bf ConnectedComponent\underline{s}}$ sequentially determines the connected compo\-nents $C_i$ of the graph $G$ by calling the procedure $\proc{\bf ConnectedCom-}$ $\proc{\bf ponent}$. $K$ denotes the number of connected components of $G$.

\begin{codebox}
\Procname{$\proc{\bf ConnectedComponent\underline{s} ($G$)};$}\\
\li $V\leftarrow V(G)$; 
\li $K\leftarrow 1$;
\li \While $V\neq\varnothing$:
 \Do
\li\mbox{select} $i\!\in\!V$;
\li $V\leftarrow V\setminus\{i\}$;
\li$C_K\leftarrow\proc{\bf ConnectedComponent ($G, i$)}$; 
\li $K\leftarrow K+1$;							
		\End\\

\li \mbox{\bf result}$\ \leftarrow \{C_1,\ldots,C_K\}$. 				
\end{codebox}

\section{The determinant of the graph matrix \\ and the permutations implemented on the graph}

\subsection{Permutations implemented on the graph} 

By definition
\begin{equation}\label{5}
\det A(G)=\sum\limits_{\pi\in S_n}(-1)^{\sigma(\pi)}a_{1\pi(1)}\cdot\ldots\cdot a_{n\pi(n)},
\end{equation}
where $S_n$ is the symmetric permutation group, $\sigma(\pi)$ is the signature of the permutation $\pi\!\in\!S_n$. A permutation $\pi\!\in\!S_n$ is said to be {\it implemented on the graph} $G$ if $\pi(i)\!=\!j$ only if there is an oriented edge $(i,j)\!\in\!E(G)$ for all $i\!=\!\overline{1,n}$. Permutation $\pi\!\in\!S_n$ is implemented on a graph $G$ with the matrix $A$ if and only if $$\prod\limits_{i=1}^na_{i\pi(i)}>0.$$ Since every permutation is a product of cycles:
\begin{equation}\label{6}
\pi=(i_1\ldots i_{l_1})(i_{l_1+1}\ldots i_{l_1+l_2})\ldots(i_{l_1+\ldots+l_{k-1}+1}\ldots i_{l_1+\ldots+l_c}),
\end{equation} where $c$ is the number of cycles in $\pi$, $l_i$ is the length of the $i^{th}$ cycle, then the permutation is implemented on the graph only if the cycle in its representation (\ref{6}) corresponds to an oriented cycle in the graph. The loop $(i,i)$ corresponds to the cycle $(i)$ in the representation $(\ref{6})$ of a permutation implemented on the graph.

Let us denote as $\mathcal{R}(G)$ the set of permutations that are implemented in the graph $G$. We have
\begin{equation}\label{7}  
\det A(G)=\sum\limits_{\pi\in\mathcal{R}(G)}(-1)^{\sigma(\pi)}a_{1\pi(1)}\cdot\ldots\cdot a_{n\pi(n)}.
\end{equation}  
Actually, since $$\det A(G)=\sum\limits_{\pi\in S_n\setminus \mathcal{R}(G)}(-1)^{\sigma(\pi)}a_{1\pi(1)}\cdot\ldots\cdot a_{n\pi(n)}+\sum\limits_{\pi\in \mathcal{R}(G)}(-1)^{\sigma(\pi)}a_{1\pi(1)}\cdot\ldots\cdot a_{n\pi(n)},$$ then, if $\pi\!\in\!S_n\setminus \mathcal{R}(G)$, there is such a vertex $i\!\in\! V(G)$ that $a_{i\pi(i)}\!=\!0$. Therefore, $$\sum\limits_{\pi\in S_n\setminus\mathcal{R}(G)}(-1)^{\sigma(\pi)}a_{1\pi(1)}\cdot\ldots\cdot a_{n\pi(n)}=0,$$ so (\ref{7}) is true.\\

\subsection{The determinant of the graph matrix as a polynomial of $d$} 

Let $C_l(G)$ be permutations of $\mathcal{R}(G)$ with the total length of cycles included in the permutation equal to $l$. Let
\begin{equation}\label{8}
c_l=c_l(G)=\sum\limits_{\pi\in C_l(G)}(-1)^{\sigma(\pi)},
\end{equation} 
that is, $c_l$ is the number of permutations with a total length equal to $l$, calculated with taking into account their parity. Then
\begin{equation}\label{9}
\det A(d)=d^n+c_2d^{n-2}+c_3d^{n-3}+\ldots+c_{n-1}d+c_n.
\end{equation} 
That is, $\det A(G)\!$ is a polynomial of $d$ from (\ref{1}). 

The coefficient $c_0(G)$ of the polynomyal for any $G$ is equal to 1, since there is the only one permutation of $S_n$ that has the total length of the cycles included in it equal to zero. It is the permutation $(1)(2)\ldots(n)$. The number of cycles of length $1$ in the graph is zero by definition of the cycle, hence $c_1(G)\!=\!0$ for any $G$. The number of cycles of length $2$ in the graph $G$ is equal to the number of its undirected edges --- the only possible cycles of length $2$ in the graph, that is, $c_2(G)\!=\!-|E(G)|$, because permutations from the set $C_2(G)$ are odd we have the minus sign. 

So, we consider $\det A(G)\!$ not only as the real values determinant of the graph matrix but we also consider it as a polynomial $\det A(d)\!$.\\

\section{Justification of the approach}

To theoretically substantiate our approach to finding the connected components of a graph, it is necessary to show that the inequality of components at step 5 of the procedure $\proc{\bf ConnectedComponent}$ is indeed a necessary and sufficient condition for the vertices $i$ and $j$ to belong to one connected component.

\bigskip

\noindent {\bf Statement 1.} {\it Let $i$ is the perturbed vertex of the graph $G$. For almost all $d$ defined by (\ref{2}), $x_j'\!\neq\!x_j$ if and only if the vertices $i$ and $j$ of the graph belong to the same connected component of $G$.}\\

\smallskip

\noindent{\bf Proof.} Let us estimate the difference of the $j^{th}$ components of the vectors $x$ and $x'$. Let $A_{ij}$ denote the submatrix of $A$ obtained from it by removing its $i^{th}$ row and $j^{th}$ column. $A'$ and $A_{ij}'$ respectively, are the matrix of the graph $G$ and its submatrix after the perturbation has been performed. We have $$\bigl |x_j'-x_j\bigr|=\biggl |\frac{\det A_{ij}'}{\det A'}-\frac{\det A_{ij}}{\det A}\biggr |.$$ Since $\det A'\!=\!\det A+\varepsilon \det A_{ii}$ and $A_{ij}'\!=\!A_{ij}$, we have
\begin{equation}\label{10}  
\bigl |x_j'-x_j\bigr |=\biggl |\frac{\det A_{ij}}{\det A+\varepsilon\det A_{ii}}-\frac{\det A_{ij}}{\det A}\biggr |=\frac{\varepsilon \det A_{ii}\,|\det A_{ij}|}{\det A(\det A+\varepsilon \det A_{ii})}.\end{equation}
$\varepsilon\!>\!0$, $\det A$ and $\det A_{ii}$ are determinants of matrices with strict diagonal predominance, so they are positive according to Hadamard's criterion. As follows from (\ref{10}), in order to show that $x_j'\!\neq\!x_j$, it is necessary to show that $\det A_{ij}\!\neq\!0$.

If the perturbed vertex $i$ and the vertex $j$ belong to the same connected component, then the value of $\det A_{ij}$ is the sum of the form (\ref{7}) of terms corresponding to permutations implemented in the oriented graph $G_{ij}$ on $n$ vertices obtained from $G$ as follows. Considering $G$ as a directed graph in which $(i,j)\!\in\! E(G)$ if and only if $(j,i)\!\in\! E(G)$, we get $G_{ij}$ from it by removing all oriented edges originating from the vertex $i$ and also removing all oriented edges entering the vertex $j$, and setting the edge $(i,j)\!\in\!V(G_{ij})$. In the $i^{th}$ row and in the $j^{th}$ column of the matrix $A(G_{ij})$, all elements are null except for the $ij^{th}$ element, which is equal to $1$. $\det A(G_{ij})=\det A_{ij}$.

Any permutation of $\mathcal{R}(G_{ij})$, due to the reachability from the vertex $i$ of the vertex $j$ in the graph $G$, contains a cycle which consists of the oriented edge $(i,j)$ and some chain of oriented edges starting from the vertex $j$ and ending at the vertex $i$. Hence, $\mathcal{R}(G_{ij})\!\neq\!\varnothing$, which means $\det A_{ij}(d)$ is some nonzero polynomial from $d$ of degree not exceeding $n\!-\!2$. So $\det A_{ij}(d)$ has no more than $n\!-\!2$ roots over $\mathbb{R}$, and hence the equality $\det A_{ij}(d)\!=\!0$ holds only on the set of values of $d$ which has measure zero. Hence, from (\ref{10}) we have: $$\bigl|x_j'-x_j\bigr |>0$$ for almost all values of $d\!\in\!\mathbb{R}$.

If $i$ and $j$ belong to different connected components, there is no chain of oriented edges starting at $j$ and ending at $i$ in graph $G_{ij}$. Hence, $\mathcal{R}(G_{ij})\!=\!\varnothing$. So $\det A_{ij}(d)\!=\!0$ regardless of the value of $d$, and it follows from (\ref{10}) that $$\bigl|x_j'-x_j\bigr|=0$$ for any $d$. $\qed$\\

\subsection{Lower bound of the value $|x_j'\!-\!x_j|$} 

It follows from (\ref{10}) that $|x_j'\!-\!x_j|\!>\!0$ if and only if the perturbed vertex $i$ and the vertex $j$ belong to the same connected component. However, in order to guarantee the fulfillment of the inequality  $x_j'\!\neq\!x_j$ for machine numbers representing the exact values of the components of the vectors $x'$ and $x$, it is necessary to estimate the difference between these values, that is, find such a minimum lower bound $\Delta\!>\!0$ that $$\bigl |x_j'-x_j\bigr |\ge\Delta.$$ 

Let $l(i,j)$ be the length of the shortest chain connecting $i,j\in V(G)$, and let $l(i,j)\!=\!\infty$ if $i$ and $j$ belong to different connected components. Let $\ell$ denote the {\it diameter} of the graph, which we define as $$\ell\!=\!\max\{l(i,j)\ | \ (i,j\!\in\!V(G))\mbox{\ and\ }(l(i,j)\!\neq\!\infty)\}.$$

\bigskip 

\noindent{\bf Lemma 1.} {\it Let $G$ be a connected graph, $i\!\in\!V(G)$ is the perturbed vertex. Then} 
\begin{equation}\label{11}
\bigl |\det A_{ij}\bigr |\!\ge\!1
\end{equation}
{\it for any $j\!\in\!V(G)$.}

\bigskip

\noindent{\bf Proof.} Let $G$ be a chain on $n$ vertices. For $i,j\!\in\!V(G)$, the minimum cardinality of $\mathcal{R}(G_{ij})$ will be reached when $l(i,j)\!=\!\ell\!=\!n$, because in this case $\mathcal{R}(G_{ij})$ consists of only one cycle: $\mathcal{R}(G_{ij})\!=\!\{\pi\}$, where $\pi\!=\!(i_1i_2i_3\ldots i_n)$ and the following conditions are met:
\begin{itemize}
\item[1).] $i_1\!=\!i$, $i_2\!=\!j$, what corresponds to an oriented edge $(i,j)\!\in\!E(G_{ij})$. 
\item[2).] $i_3,\ldots,i_n$ are vertices corresponding to the chain from vertex $j$ to vertex $i\!=\!i_n$ consisting of oriented edges $(i_k,i_{k+1})\!\in\!E(G_{ij})$. Such a chain exists because, by the lemma condition, the vertices $i$ and $j$ belong to the same connected component.
\end{itemize}
\noindent Therefore, by (\ref{7}) we have 
\begin{equation}\label{12}
\det A_{ij}=(-1)^{\sigma(\pi)}\cdot 1,
\end{equation}
that is, $\bigl |\det A_{ij}\bigr|=1$.

For any connected graph $G$ obtained from a simple chain on $n$ vertices by adding edges to it, the power of $\mathcal{R}(G)$ will not decrease, which means the number of terms in (\ref{7}) will also not decrease. Thus, $\det A_{ij}\!=\!\det A_{ij}(d)$ will be a nonzero polynomial of the form (\ref{9}) defined for the directed graph $G_{ij}$. Therefore, for sufficiently large $d$ we have $\bigl|\det A_{ij}(d)\bigr|\ge 1$ for arbitrary connected undirected graph $G$ with matrix $A$. $\qed$

\bigskip

\noindent{\bf Lemma 2.} {\it Let $G$ be a connected graph, $i\!\in\!V(G)$ is the perturbed vertex, $\varepsilon\!=\!1$. Then} 
\begin{equation}\label{13}
|x_j'-x_j|\ge\frac{1}{2d^{2n}}
\end{equation}
{\it for any vertex} $j\!\in\!V(G)$.
\bigskip

\noindent{\bf Proof.} Consider the estimate (\ref{10}). Since $A_{ii}$ is a matrix with strict diagonal predominance, so the Hadamard's conditions are met for it and, since $d\!=\!\mu\cdot d_{\max}$, $\mu\!>\!1$, we have
\begin{equation}\label{14}
\det A_{ii}\ge\prod\limits_{i=1}^{n-1}(d-d_i)>1.
\end{equation} Since $$\det A\le\biggl ( \frac{1}{n}\ \mbox{tr} A\biggr )^n,\quad\det A_{ii}\le\biggl ( \frac{1}{n-1}\ \mbox{tr} A_{ii}\biggr )^{n-1},$$ and $\mbox{tr} A=nd$, $\mbox{tr} A_{ii}=(n-1)d$, then 
\begin{equation}\label{15}
\det A\le\biggl ( \frac{1}{n}\ n d\biggr )^n=d^n,
\end{equation}
\begin{equation}\label{16}
\det A_{ii}\le\biggl ( \frac{1}{n-1}\ (n-1) d\biggr )^{n-1}=d^{n-1}.
\end{equation}
Since $|\det A_{ij}|\!\ge\! 1$ by Lemma 1, then from (\ref{10}) and (\ref{14}) -- (\ref{16}) we have:
$$|x_j'-x_j|\ge\frac{1}{d^n(d^n+d^{n-1})}\ge\frac{1}{2d^{2n}}.$$ $\qed$

\subsection{The required accuracy needed to implement tha approach numerically. Computational complexity of the approach}

To implement our approach we need to solve SLAEs of the form $(\ref{4})$ with required accuracy. We can use iterative methods for SLAEs such as the simple iteration method (Jacobi method) or the Gauss-Seidel method. The diagonal predominance given by (\ref{1}) ensures the convergence of these methods to the exact solution at a geometric progression rate. In addition, the use of these methods is convenient for us, since it easily allows us to estimate the required computational complexity of finding the solution of the SLAE with the accuracy that is necessary for the implementation of the approach.

Let the Gauss-Seidel method be used to solve SLAE $(\ref{4})$, and let $x^{(k)}$ is an approximation of the exact solution of $x$ on the $k^{th}$ iteration of the method. We have: $$\alpha=\max\limits_{1\le i\le n}\frac{\sum\limits_{i\neq j}|a_{ij}|}{|a_{ii}|}<1,$$ since the matrix $A$ of the form (\ref{1}) has a strict dagonal predominance. Consequently, the Gauss-Seidel method converges to an exact solution at the rate of geometric progression \cite{berezin}: $$\bigl\|x-x^{(k)}\bigr\|_1\le\alpha\bigl\|x-x^{(k-1)}\bigr\|_1,$$ that is equivalent to $$\bigl\|x-x^{(k)}\bigr\|_1\le\alpha^k\bigl\|x-x^{(0)}\bigr\|_1.$$ For matrices of the form (\ref{1}) for which $d\!=\!\mu\cdot d_{\max}$, $\mu\!>\!1$, we have: $$\alpha\le\frac{d_{\max}}{d}\le\frac{d_{\max}}{\mu\cdot d_{\max}}=\frac{1}{\mu}.$$ Therefore, when solving SLAE (\ref{4}) by the Gauss-Seidel method, we have: $$\bigl\|x-x^{(k)}\bigr\|_1\le\frac{1}{\mu^k}\bigl\|x-x^{(0)}\bigr\|_1.$$ It means that $$\bigl |x_j-x_j^{(k)}\bigr|\le\bigl\|x-x^{(k-1)}\bigr\|_1\le\frac{\delta_0}{\mu^k},$$ where $\delta_0$ is the error of the initial approximation.

\smallskip

Consider the following problem. Let $a$, $b\!\in\!\mathbb{R}$, and $a^{(k)}$, $b^{(k)}\!\in\!\mathbb{R}$ are their approximations obtained for $k$ of the first iterations of the numerical method such that $$\bigl|a-a^{(k)}\bigr |\le\frac{\delta_0}{\mu^k},\qquad\bigl|b-b^{(k)}\bigr|\le\frac{\delta_0}{\mu^k}.$$

Suppose we know such $\Delta\!>\!0$ that if $a\!\neq\!b$, then $|a-b|\!\ge\!\Delta$. Let's determine how many iterations of the numerical method will be required in order to testify the inequality of the exact values of $a,b\!\in\!\mathbb{R}$ relying on the  inequality of machine numbers $a^{(k)}$ and $b^{(k)}$ that represent them, $a^{(k)}, b^{(k)}\in\mathbb{Q}$.

The fulfillment of the inequalities $$\bigl|a-a^{(k)}\bigr |<\frac{\Delta}{4},\qquad\bigl|b-b^{(k)}\bigr |<\frac{\Delta}{4}$$ will be followed by the fulfillment of the inequality $$\bigl |a^{(k)}-b^{(k)}\bigr |>\frac{\Delta}{2},$$ and on subsequent iterations of the numerical method the difference between approximate values of $a^{(k)}$ and $b^{(k)}$ will only grow. So, it can be argued that $a\!\neq\!b$.

\smallskip

Thus, if $|a-b|\!\ge\!\Delta\!>\!0$, then, using machine numbers with a mantissa length that is sufficient to fix the inequality of the exact values of $a\!\neq\!b$, it is necessary to carry out such a number $N$ of iterations of the numerical method that $$\frac{\delta_0}{\mu^N}<\frac{\Delta}{4}.$$ Since $\|x\|_1\le\|A^{-1}\|_1\|e_i\|_1\le 1$, then, choosing the initial approximation of $x^{(0)}\!=\! e_i$ for determining the connected component containing the vertex $i$, we have $\delta_0\!<\!1$, which means $\log_{\mu}\delta_0\!<\!0$. That is, it is necessary that the inequality $$N>\log_{\mu}\biggl (\frac{4\delta_0}{\Delta}\biggr )$$ be fulfilled. Since by Lemma 2 we have $|x_j'-x_j|\geq\Delta$, where $\Delta\!=\!1/(2d^{2n})$, then $$N\ge 2n\log_{\mu}d+\log_{\mu}(8\delta_0)=2n\log_{\mu}d+\log_{\mu}8+\log_{\mu}\delta_0.$$ Therefore, in order to achieve the required accuracy, it is sufficient to produce $N$ iterations, where $$N\ge 2n\log_{\mu}d+1.$$ Since $$\log_{\mu}d=\log_{\mu}(\mu d_{\max})=1+\log_{\mu}d_{\max},$$ then with $\mu\!=\!d_{\max}$ we have: $$N\ge 4n+1,$$ i.e. $N\!=\!O(n)$. We have proved

\bigskip

\noindent{\bf Statement 2.} {\it It is enough to produce $O(n)$ iterations of the Gauss-Seidel method to achieve the required accuracy of solution of the SLAE (\ref{4}) in order to implement the algorithm $\proc{\bf ConnectedComponent\underline{s}}$ numerically.}\\

\subsection{Required mantissa length of machine numbers} 

To substantiate the possibility of numerical implementation of the proposed approach, it is also necessary to show that the inequality $x_j'\!\neq\!x_j$ can be fixed numerically using machine numbers with a mantissa length is at lest a function of polynomial growth of $n$.

In order to fix a non-zero value of $|x_j'-x_j|$ that is bounded from below by the value of $\Delta\!=\!1/(2d^{2n})$, it requires such a mantissa length $L$ that $$\frac{1}{10^L}\le\frac{1}{2d^{2n}},$$ that is $$10^L\ge 2d^{2n}.$$ Therefore, $$L\ge\lg d^{2n}+\lg 2.$$ That is, $L$ must satisfy the inequality $$L\ge 2n\lg d+1=2n(\lg\mu +\lg d_{\max}).$$ When choosing $\mu\!=\!d_{\max}$, we have $L\!=\!O(n)$. Thus we have shown that required machine numbers mantissa length is a function of linear growth of $n$.\\

\section{Data structures and procedures that accelerate algorithms for finding graph connectivity components}

\subsection{Calculations with a matrix portrait}

By a {\it portrait} of the matrix $A(G)$ we mean a one-dimensional, not necessarily ordered array $\mbox{Ap}$ containing $m\!=\!E(G)$ elements $\mbox{Ap[}t\mbox{]}$, $t\!=\overline{1,m}$. Each element of $\mbox{Ap[}t\mbox{]}$ contains two values: the values $\mbox{Ap[}t\mbox{].v1}$ and $\mbox{Ap[}t\mbox{].v2}$ that correspond to edges of $G$. That is, $$\bigl (\forall(i,j)\!\in\!E(G)\bigr) \bigl (\exists t : 1\le t\le m\bigr ) \bigl ((\mbox{Ap[}t\mbox{].v1}\!=\!i) \mbox{ and } (\mbox{Ap[}t\mbox{].v2}\!=\!j)\bigr ).$$

Let us consider the computations using the method of simple iteration when working with the portrait of the adjacency matrix. One iteration of the numerical method is performed in one passage through the portrait. To get the component $x_i^{(k+1)}$ of the approximate solution $x^{(k+1)}$ we sum such components $x_j^{(k)}$ of the previous approximate solution $x^{(k)}$, that $\mbox{Ap}$ contains an element corresponding to $(i,j)\!\in\!E(G)$.

\begin{codebox}
\Procname{$\proc{\bf SimpleIteration}\ (x^{(k)}):\ x^{(k+1)};$}\\

\li \For $t=1$ \To $m$:
 \Do
\li $i\leftarrow \mbox{Ap[}t\mbox{].v1}$, $j\leftarrow\mbox{Ap[}t\mbox{].v2}$;
\li $x_i^{(k+1)}\leftarrow x_i^{(k+1)}+x_j^{(k)}$;
\li $x_j^{(k+1)}\leftarrow x_j^{(k+1)}+x_i^{(k)}$; 
 \End\\
 
\li \For $i=1$ \To $n$:
 \Do
\li\If $i$ is the {\mbox perturbed vertex}
 \Then
\li $x_i^{(k+1)}\leftarrow\bigl (1-x_i^{(k+1)}\bigr )/d;$
\li \Else
\li $x_i^{(k+1)}\leftarrow\bigl (0-x_i^{(k+1)}\bigr )/d;$ 
 \End\\
 
\li \mbox{\bf result}\ $\leftarrow x^{(k+1)}$. 
\end{codebox}

Compuatations with a portrait of a matrix, implemented by the procedure, takes time of $O(m+n)$. For sparse graphs, the computationanl complexity of the procedure is $O(m)$. 

Iterations of the Gauss-Seidel method were similarly organized by us as computations with a portrait of the graph's matrix.\\

\subsection{Masking of vertices}

To speed up computations of algebraic BFS, masking of vertices is used. {\it Masking of vertices} of a graph $G$ means marking 
rows, columns of matrix $A(G)$ and components of vectors $x^{(k)}$ with numbers corresponding to vertices from some subset of $V(G)$ that we used to ignore during computations. Thus, for example, if we have already find some connected component $V_k$ then we may ignore computations with $x_j^{(k)}$ such that $j\!\in\! V_k$ performing them only for unmasked part of the vector. This is equivalent to reducing the dimension of the problem being solved.

So, if $|V(G)|\!=\!n$ and a connected component with a set of vertices $V_k$ is obtained at the iteration of the algorithm $\proc{\bf ConnectedComponent\underline{s}}$,  $|V_t|\!=\!n_t$, then in subsequent iterations of solving SLAE in procedure $\proc{\bf ConnectedComponent}$, by multiplying the vector on the matrix $A$, we can only work with those rows and columns of $A$ and with those components of aprroximate solutions that correspond to the vertices of $V(G)\setminus V_t$. As a result, we get a reduction in the dimension of the problem being solved by $n_k$. 

Different ways of masking of vertices may be used. Despite the fact that the implementation of masking itself requires computational costs, masking can be very effective. It is the implementation of the specific masking of the {\it visited} vertices that allows to obtain for sparse graphs an algorithm for finding the connected components with linear computational complexity $O(m)$, i.e. with the minimum possible complexity for the sequential implementation of the algebraic BFS \cite{Burkhardt}.

Masking can also be applied to the approach with perturbations of the graph matrix. We can do the same for modifications of procedure $\proc{\bf ConnectedCompo-}$ $\proc{\bf nent}$, discussed below.\\

\section{Computational complexity of the approach}

The computational complexity of the approach we present is determined by the computational complexity of solving SLAE (\ref{4}) on iterations of the algorithm $\proc{\bf ConnectedComponent}$. According to Statement 2, $O(n)$ iterations of the numerical method are required to obtain an approximate solution with sufficient accuracy. Using the portrait of the graph matrix, it takes $O(m)$ of elementary machine opertaions to perform one iteration of the algorithm. Thus, the computational complexity of finding one connected component is $O(mn)$. So, for the bounded number of connected components $K$ the total complexity of the algorithm $\proc{\bf ConnectedComponent\underline{s}}$ is equal to $O(mn)$.

At each iteration of the algorithm $\proc{\bf ConnectedComponent}$ it is required to store SLAE's matrix and vectors $x^{(k)}$ and $x^{(k+1)}$. Using matrix portret, it requires $O(m\!+\!n)$ memory cells.

\smallskip

Our approach allows to use numerical methods of solving SLAE to solve connectivity problems on graphs. The algorithm $\proc{\bf ConnectedComponent\underline{s}}$ is equivalent in its computational complexity to algebraic BFS. It can be implemented using efficient parallel implementations of numerical SLAE solution methods and, in particular, their implementations for sparse graphs. Also, instead of the implementations of iterative methods for solving SLAE in the algorithm $\proc{\bf ConnectComponent}$, effective parallel implementations of non-iterative numerical methods for solving SLAE can also be used.\\

\section{Iterative numerical methods for solving SLAE\\ and graph traversals}

\subsection{Iterative numerical methods for solving SLAE\\ as implementations of graph traversals}

For the algorithm $\proc{\bf ConnectedComponent\underline{s}}$, when computing the components of the vectors $x$ and $x'$, it does not matter which numerical method of solving SLAE we shall use. It is necessary that using this method we obtain an approximate solution so close to the exact one that different exact solutions of two SLAEs are distinguishable in the representation of their components by machine numbers with a fixed mantissa length. In the above discussion, iterative methods were convenient for us, because it is convenient for us to obtain estimates of the computational complexity of the SLAE solution with a given accuracy.

However, if a relatively small number of iterations of the iterative methods of solving SLAE are performed without achieving convergence to the exact solution, then different iterative numerical methods will set different sequences of approximate solutions that will correspond to different traversals of the graph. So, for example, as we will show later, the traversal that sets the simple iteration method will be equivalent to the traversal that implements BFS, whereas the traversal that sets the Gauss-Seidel method is not equivalent to it.

\smallskip

{\it Graph traversal} is the process of visiting (checking and/or updating) each vertex in a graph. Graph traversal, which is performed during the iterative numerical method, is defined as follows: at the $k^{th}$ iteration of the numerical method, there is a transition to the $j^{th}$ vertex from one of the previously reached vertices, if
\begin{equation}\label{17}
(x_j^{(k)}\!=\!0)\mbox{ and }(x_j^{(k+1)}\!\neq\! 0).
\end{equation}
$x_j^{(k)}\!=\!0$, if the vertex $j$ is not reached yet by the $k^{th}$ iteration during the traversal, which is given by the numerical method, and $x_j^{(k)}\!\neq\!0$ otherwise.

A {\it chain with the correct order coming from the vertex} $j$ is such a simple chain $\{j_1,j_2,\ldots,j_s\}$, $j_t\!\in\!V(G)$, $s\!\ge\!1$, that $j_1\!=\!j$, $j_t\!\le\! j_{t+1}$, $t\!=\!\overline{1,s\!-\!1}$. 

\smallskip

Let us consider the traversals that are implemented by the simple iteration and Gauss-Seidel methods. We assume $x^{(0)}\!=\! b\!=\!e_i$ for both. Iterations of the simple iteration method have the form:
\begin{equation}\label{18}
x^{(k+1)}=b-D^{-1}Ax^{(k)},
\end{equation}
i.e.  
\begin{equation}\label{19}
x_j^{(k+1)}=\frac{1}{a_{jj}}\biggl (b_j-\sum\limits_{l\neq j}a_{jl}x_j^{(k)}\biggr ).
\end{equation}
In the sense of the definition given above, the traversal implemented by the simple iteration method is equivalent to the traversal implemented by algebraic BFS:
\begin{equation}\label{20}
x^{(k+1)}=Ax^{(k)},
\end{equation} 
since for sequences of vectors (\ref{18}) and (\ref{20}), the condition (\ref{17}) will be met for the same values of $k$.

The transformation of the vector $x^{(k)}$, performed on the iteration of the Gauss-Seidel method, is given as 
\begin{equation}\label{21}
(L+D){x}^{(k+1)}=-U {x}^{(k)}+b,
\end{equation} 
i.e. 
\begin{equation}\label{22}
x_j^{(k+1)}=\frac{1}{a_{jj}}\biggl (b_j-\sum\limits_{l=1}^{j-1}a_{jl}x_l^{(k+1)}-\sum\limits_{l=j+1}^{n}a_{jl}x_l^{(k)}\biggr ).
\end{equation} 
The traversal implemented on iterations of the Gauss-Seidel method is not equivalent to the traversal implemented on BFS iterations, since for sequences of vectors (\ref{20}) and (\ref{22}), the condition (\ref{17}), generally speaking, will be met for different values of $k$. This is also true for some graphs from the examples we consider below.

\bigskip

Consider the following procedures for finding the connected component of a graph to which a given vertex belongs. During the operation of the first such procedure, modified iterations of the simple iteration method are performed. During the work of the second we perform  modified iterations of the Gauss-Seidel method. $b_j\!=\!1$ if $j\!=\!i$, and $b_j\!=\!0$ otherwise. These procedures can be used instead of the procedure $\proc{\bf ConnectedComponent}$ called by the algorithm $\proc{\bf ConnectedComponent\underline{s}}$.

\begin{codebox}
\Procname{$\proc{\bf ConnectedComponent2}\ (G, i\in\!V(G)): C;$}
\li $x^{(0)}=e_i$; $C\leftarrow\{i\}$;
\li $k\leftarrow 0$;

\li\Repeat
		   
\li    \For $\forall j\in\!V(G)$:
		     \Do
\li		    $x_j^{(k+1)}=a_{jj}\cdot\biggl (b_j-\sum\limits_{l=1}^{n}a_{jl}x_l^{(k)}\biggr ).$			
         \End 
\li      $k\leftarrow k+1$;

\li\Until $\bigl (\exists j\!\in\!V(G)$ : $( x_j^{(k)}=0)\mbox{ and }(x_j^{(k+1)}\neq 0) \bigr )$			

\li $C\leftarrow C\cup \{j\!\, :\!\, x_j^{(k)}\neq 0\}$;\\

\li \mbox{\bf result} $\ \leftarrow C$. 		

\end{codebox}
	
\bigskip

\begin{codebox}
\Procname{$\proc{\bf ConnectedComponent3}\ (G, i\in\!V(G)): C;$}
\li $x^{(0)}=e_i$; $C\leftarrow\{i\}$;
\li $k\leftarrow 0$;

\li\Repeat
\li    \For $\forall j\in\!V(G)$:
		     \Do
\li		    $x_j^{(k+1)}=a_{jj}\cdot\biggl (b_j-\sum\limits_{l=1}^{j-1}a_{jl}x_l^{(k+1)}-\sum\limits_{l=j+1}^{n}a_{jl}x_l^{(k)}\biggr ).$			
         \End 
\li      $k\leftarrow k+1$;

\li\Until $\bigl (\exists j\!\in\!V(G)$ : $( x_j^{(k)}=0)\mbox{ and }(x_j^{(k+1)}\neq 0) \bigr )$			
		
\li $C\leftarrow C\cup \{j\!\, :\!\, x_j^{(k)}\neq 0\}$;\\

\bigskip

\li \mbox{\bf result}$\ \leftarrow C$. 		

\end{codebox}

\bigskip

The difference between the iterations produced during these procedures from the original iterations of the simple iteration and Gauss-Seidel methods is as follows:
\begin{itemize}
\item[1)] the diagonal predominance set for matrices of the form (1) is not assumed, it is only necessary that the diagonal elements be nonzero;
\item[2)] instead of the diagonal element division operation used in (\ref{19}) and (\ref{22}), in the course of iterations $\proc{\bf ConnectedComponent2}$ and $\proc{\bf ConnectedCom-}$ $\proc{\bf ponent3}$ (step 5 of the procedures), multiplication by the element is computationally more efficient, since machine implementations of the division operation on average require three times more elementary machine operations than multiplication.
\end{itemize}

\smallskip

The algorithm $\proc{\bf ConnectedComponent\underline{s}}$, which uses the procedure $\proc{\bf Con-}$ $\proc{\bf nectedComponent2}$ to find the connected component to which a given vertex belongs, will be further called SIS abbreviated from {\it Simple Iterate Search}, and the one that uses the procedure $\proc{\bf ConnectedComponent3}$ we shall call futher as GSS abbreviated from {\it Gauss-Seidel Search}.

\smallskip

The difference between the traversal implemented by GSS and the traversal implemented by BFS and SIS is as follows. At some iteration of GSS, if a vertex is reached from which chains with the correct order originate, then at the same iteration of the procedure $\proc{\bf ConnectedComponent3}$, a traversal will be performed over all of the vertices of this chain. That is, comparing to the vertices belonging to the level of the BFS reachability tree, the vertices that are at a greater distance from the starting vertex $i$ will be reached by the same iteration of GSS. 

If the propagation of a perturbation from the vertex $i$ to the vertex $j$ is understood to mean the fulfillment of the condition (\ref{17}) for the vertex $j$ at some iteration of the procedure $\proc{\bf ConnectedComponent3}$, when does this happen, the perturbation propagates through all of the correct chains that originated from the vertices reached in the course of traversal at this iteration.\\

\subsection{Computational complexity of GSS}

The number of GSS iterations for some graph is determined by the numbering of the vertices of the graph and the choice of the starting vertex. During the execution of GSS in no more than $\ell$ iterations, all vertices belonging to the same connected component as the starting vertex will be reached, that is, the condition (\ref{17}) will be met for them. Therefore, it will take no more than $\ell$ iterations with computational complexity $O(m)$ each to find one connectivity component. Since the number of graph connected components in the problems under consideration is constant, the computational complexity of GSS is $O(m\ell)$.

Note that replacing multiplication by division at step 5 of the procedure $\proc{\bf ConnectedComponent3}$ allows us to speed up calculations by more than twice.

\smallskip

For any individual problem of finding connected components, the computational comp\-lexity of GSS does not exceed the computational complexity of algebraic BFS. The complexity of GSS is less than the complexity of algebraic BFS if, during the iterations of GSS, vertices lying on a chain with a length equal to the diameter $\ell$ of the graph are reached, and from which chains with the correct order of length from $1$ and more originate. Obviously, for most labeled undirected graphs, this condition will be met.\\

\subsection{Examples}

Consider the traversals implemented by SIS (BFS) and GSS for the graph shown in Fig. 1. The formation of $x^{(k)}$ during the iteration of the procedure $\proc{\bf ConnectedComponent2}$ called by SIS is given by the following system of equalities:
\begin{equation}\label{23}
\left\{
\begin{array}{l}
x_1^{(k+1)}=a_{11}\cdot\bigl (1-x_2^{(k)}\bigr );\\
x_2^{(k+1)}=a_{22}\cdot\bigl (-x_1^{(k)}-x_3^{(k)}-x_6^{(k)}\bigr );\\
x_3^{(k+1)}=a_{33}\cdot\bigl (-x_2^{(k)}-x_4^{(k)}-x_7^{(k)}\bigr );\\
x_4^{(k+1)}=a_{44}\cdot\bigl (-x_3^{(k)}\bigr );\\
x_5^{(k+1)}=a_{55}\cdot\bigl (-x_6^{(k)}\bigr );\\
x_6^{(k+1)}=a_{66}\cdot\bigl (-x_2^{(k)}-x_5^{(k)}-x_7^{(k)}\bigr );\\
x_7^{(k+1)}=a_{77}\cdot\bigl (-x_3^{(k)}-x_6^{(k)}-x_8^{(k)}\bigr );\\
x_8^{(k+1)}=a_{88}\cdot\bigl (-x_7^{(k)}\bigr ).\\
\end{array}\right.
\end{equation}

\begin{figure}[htbp!]
	\begin{center}
		\includegraphics[width=320px]{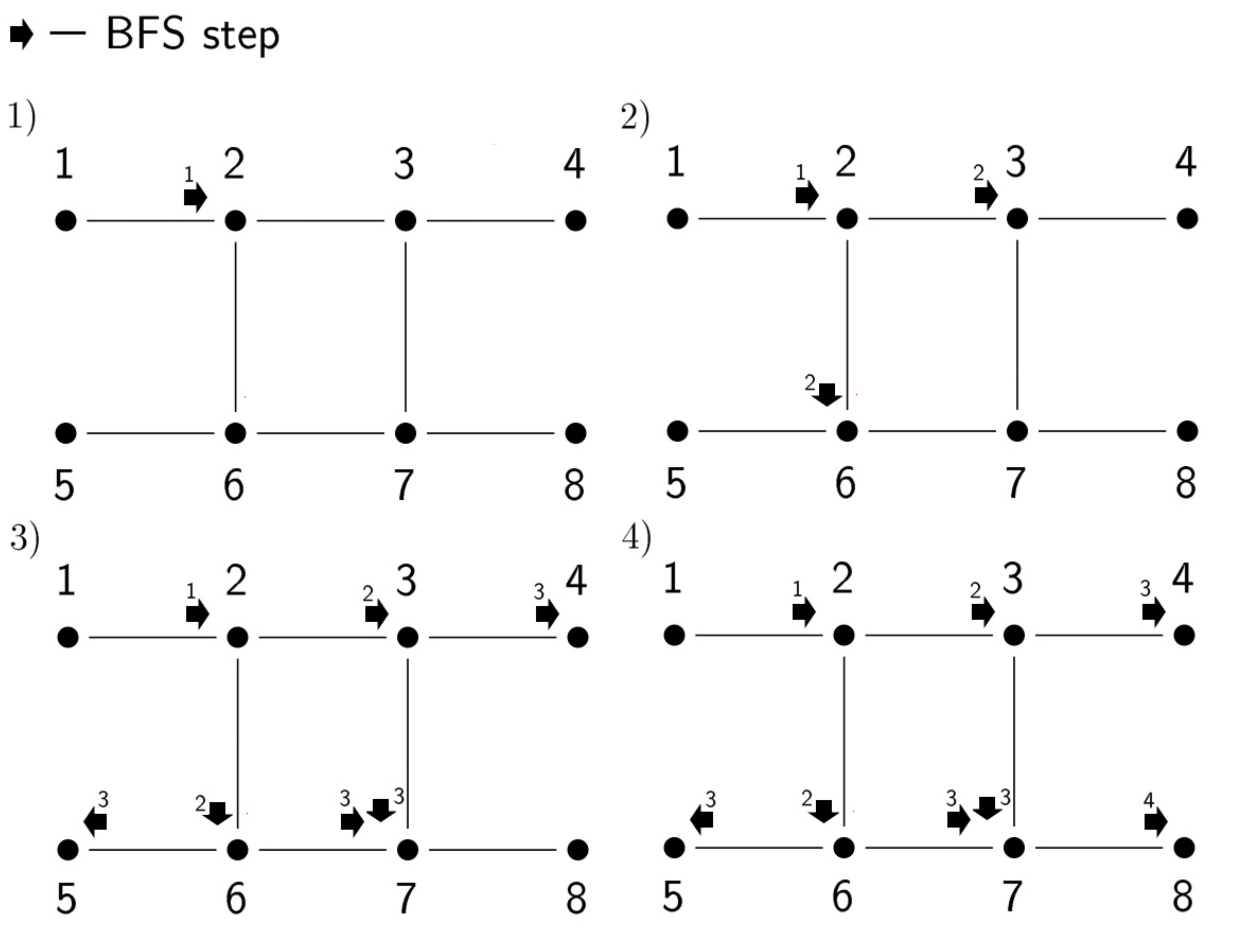}
       \caption{Traversal implemented by the procedure $\proc{\bf ConnectedComponent2}$ (SIS).}
       \label{fig1}
   \end{center}
	\end{figure}

\begin{figure}[htbp!]
	\begin{center}
		\includegraphics[width=320px]{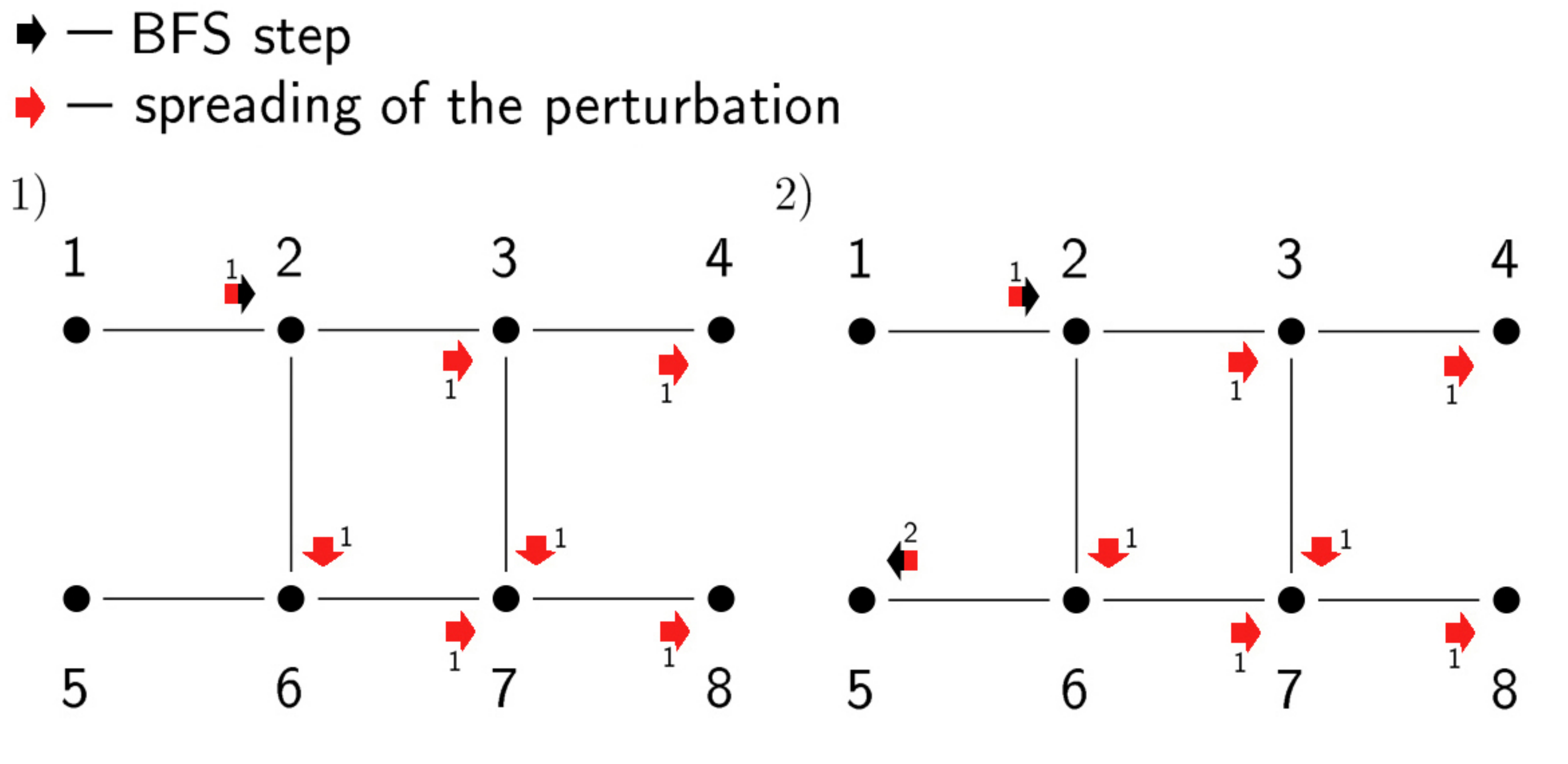}
       \caption{Traversal implemented by the procedure $\proc{\bf ConnectedComponent3}$ (GSS).}
       \label{fig2}
   \end{center}
	\end{figure}

The transformation of $x^{(k)}$ during the iteration of the procedure $\proc{\bf Connected}$ $\proc{\bf Component3}$ called by GSS is given by the equality system:
\begin{equation}\label{24}
\left\{
\begin{array}{l}
x_1^{(k+1)}=a_{11}\cdot\bigl (1-x_2^{(k)}\bigr );\\
x_2^{(k+1)}=a_{22}\cdot\bigl (-x_1^{(k+1)}-x_3^{(k)}-x_6^{(k)}\bigr );\\
x_3^{(k+1)}=a_{33}\cdot\bigl (-x_2^{(k+1)}-x_4^{(k)}-x_7^{(k)}\bigr );\\
x_4^{(k+1)}=a_{44}\cdot\bigl (-x_3^{(k+1)}\bigr );\\
x_5^{(k+1)}=a_{55}\cdot\bigl (-x_6^{(k)}\bigr );\\
x_6^{(k+1)}=a_{66}\cdot\bigl (-x_2^{(k+1)}-x_5^{(k+1)}-x_7^{(k)}\bigr );\\
x_7^{(k+1)}=a_{77}\cdot\bigl (-x_3^{(k+1)}-x_6^{(k+1)}-x_8^{(k)}\bigr );\\
x_8^{(k+1)}=a_{88}\cdot\bigl (-x_7^{(k+1)}\bigr ).\\
\end{array}\right.
\end{equation}  

By iterating (\ref{23}) with $x^{(0)}\!=\! e_1$, we get a traversal of the graph on four vertices in four iterations (Fig. \ref{pic:sis-iterations}). Unlike SIS, when iterating (\ref{24}) during GSS operation, the perturbation of vertex 1, i.e. the perturbation of the diagonal element $a_{11}$, will propagate over all chains with the correct order starting from the vertex $2$, which is reached from the vertex $1$ at the first iteration (Fig. 2). As a result, all vertices of the graph will be reached in the course of two iterations.

\begin{figure}[htbp!]
	\begin{center}
		\includegraphics[width=200px]{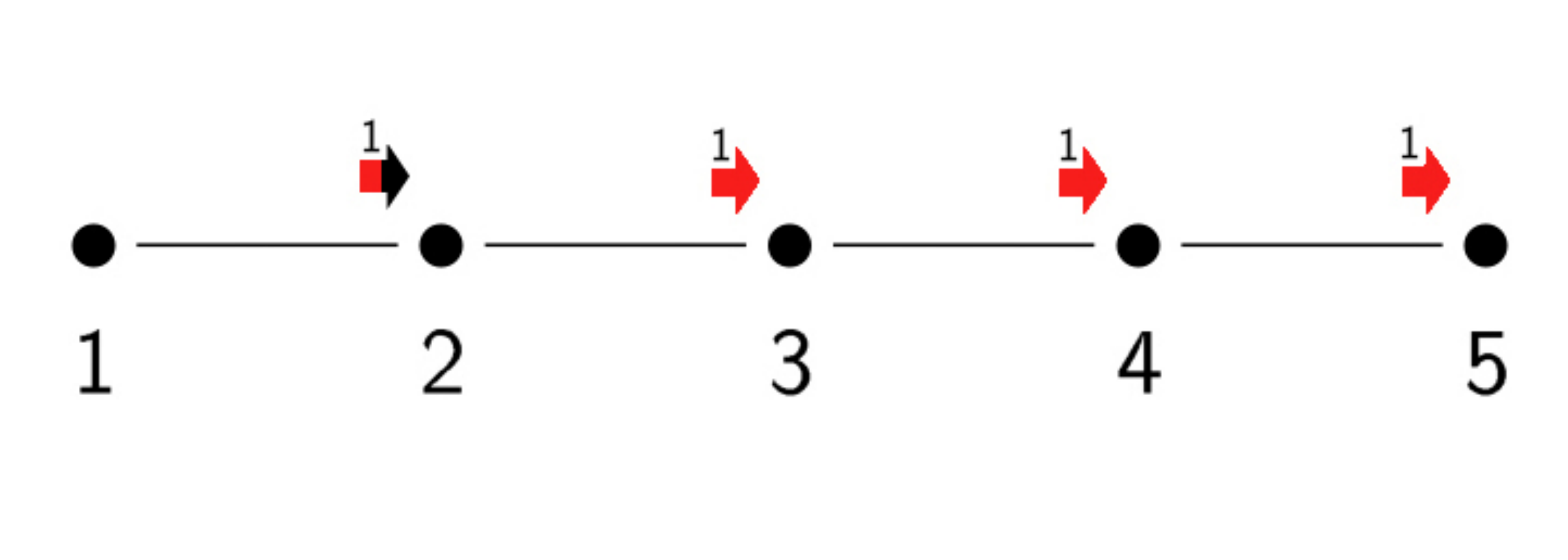}
       \caption{Traversal implemented by the procedure $\proc{\bf ConnectedComponent3}$ (GSS).}
       \label{fig3}
   \end{center}
	\end{figure}

Figures 3 and 4 show the traversals implemented by SIS and GSS for a graph representing a chain on $5$ vertices for different vertex numberings. The systems of equalities (\ref{25}) and (\ref{26}), respectively, define the SIS and GSS transformations for this graph.
\begin{equation}\label{25}
\left\{
\begin{array}{l}
x_1^{(k+1)}=a_{11}\cdot\bigl (1-x_2^{(k)}\bigr );\\
x_2^{(k+1)}=a_{22}\cdot\bigl (-x_1^{(k)}-x_3^{(k)}\bigr );\\
x_3^{(k+1)}=a_{33}\cdot\bigl (-x_2^{(k)}-x_4^{(k)}\bigr );\\
x_4^{(k+1)}=a_{44}\cdot\bigl (-x_3^{(k)}-x_5^{(k)}\bigr );\\
x_5^{(k+1)}=a_{55}\cdot\bigl (-x_4^{(k)}\bigr ).\\
\end{array}\right.
\end{equation}

\begin{equation}\label{26}
\left\{
\begin{array}{l}
x_1^{(k+1)}=a_{11}\cdot\bigl (-x_5^{(k)}\bigr );\\
x_2^{(k+1)}=a_{22}\cdot\bigl (-x_3^{(k)}\bigr );\\
x_3^{(k+1)}=a_{33}\cdot\bigl (-x_2^{(k+1)}-x_4^{(k)}\bigr );\\
x_4^{(k+1)}=a_{44}\cdot\bigl (-x_3^{(k+1)}-x_5^{(k)}\bigr );\\
x_5^{(k+1)}=a_{55}\cdot\bigl (-x_1^{(k+1)}-x_4^{(k+1)}\bigr ).\\
\end{array}\right.
\end{equation}  

For the first numbering of vertices (Fig. 3), we have the correct order for the chains originating from the vertex $2$, reached on the $1^{th}$ iteration of the both SIS and GSS algorithms. As a result, GSS will require one iteration to reach all vertices of the graph (traversing all vertices), whereas SIS will require four. Assuming that $x^{(0)}\!=\! e_1$, for the second numbering of vertices (Fig. 4) there are no chains with the correct order for all vertices reached on GSS iterations. Therefore, to reach all the vertices of the graph, both algorithms need to perform $4$ iterations, and the traversal implemented by GSS will completely repeat the traversal implemented by SIS (BFS).\\

\begin{figure}[htbp]
	\begin{center}
		\includegraphics[width=200px]{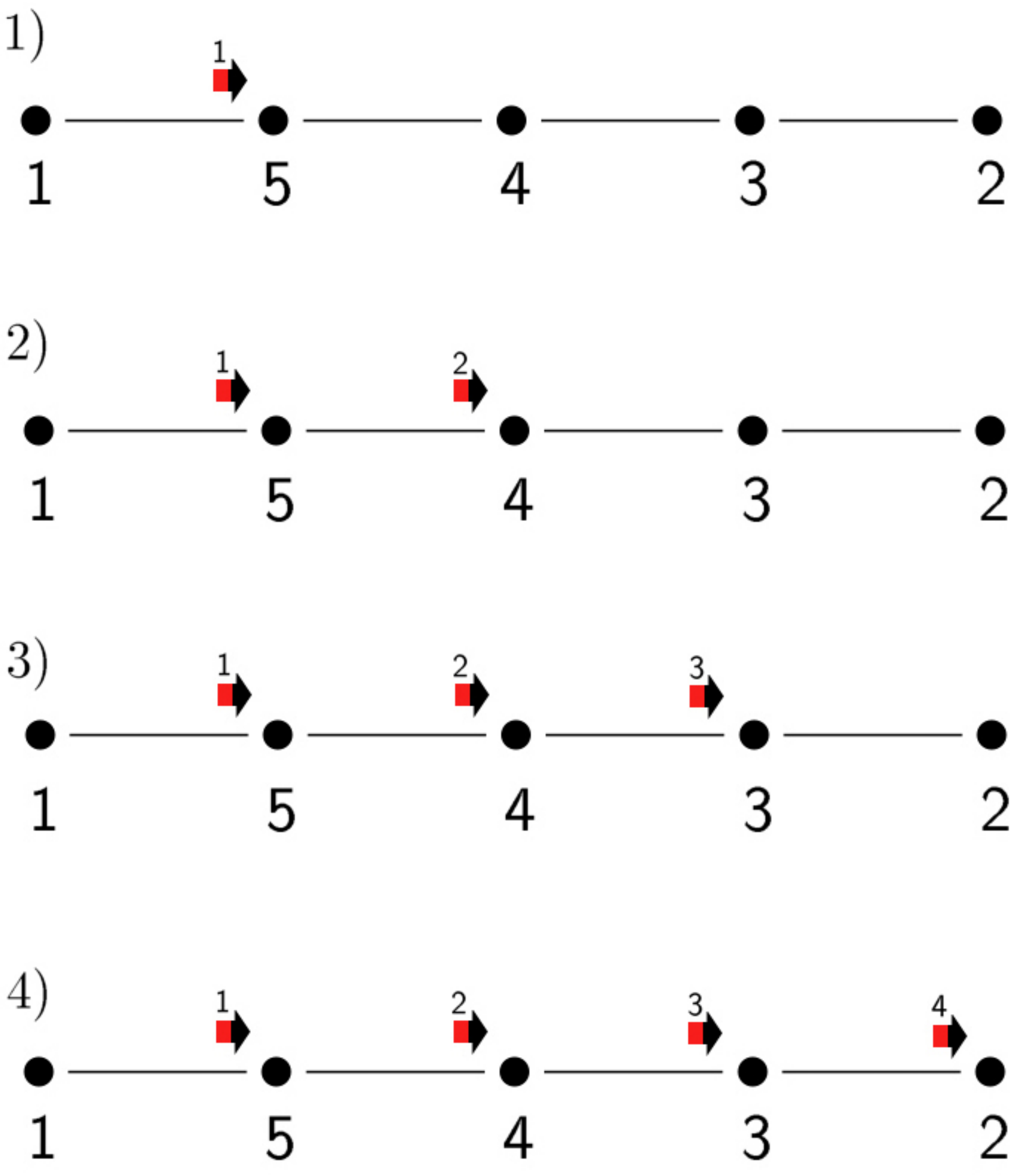}
       \caption{Traversal implemented by the procedure $\proc{\bf ConnectedComponent3}$ (GSS).}
       \label{fig4}
   \end{center}
	\end{figure}

\section{Testing algorithms}

To test the algorithms considered, we used randomly generated graphs and a sparse graph of the connectivity problem associated with  transport network of a large region of the Russian Federation.

Random sparse graphs were generated by us as graphs representing the union of chains of the same length. The numbering of the vertices of the graph was set randomly. Note that, in accordance with Lemma 1, it is for the terminal vertices of a simple chain that the lower bound in (\ref{13}) is achieved, which means that they require the maximum number of iterations to find connected components and for GSS with the worst vertex numbering for it.

The experiments were carried out on a PC with a processor clock speed of 3.8 GHz. When solving a problem with $n\!=\!90\ 000$, $ m\!=\!89\ 100$, $K\!=\!900$ using the simple iteration method, it takes about $20$ minutes to solve SLAE, while using the Gauss-Seidel method it takes $10$ minutes. At the same time, the GSS algorithm took about $3$ minutes.

In the graph of the connectivity problem associated with the transport network, the presence of an edge corresponds to the occurrence of variables in one equation or inequality of a partially integer linear programming problem. $n=367\ 840$, $m=53\ 404\ 685$, $K\!=\!224$ for the graph. It took $97$ min to find $32$ connected components containing $11\ 429$ vertices each, and $192$ connected components representing chains of $11$ vertices. Note that for a non-algebraic BFS implemented via graph traversal, it took $48$ hours to solve the same problem.\\

\section*{Conclusions}

We have proposed an approach to solving connectivity problems on graphs using perturbations of the adjacency matrix. This approach makes it possible to solve connectivity problems using effective implementations of numerical methods for solving SLAE. Such iterative methods of solving SLAE as the simple iteration method and the Gauss-Seidel method are considered by us as implementations of graph traversals. Generally speaking, the traversal is not equivalent to the traversal that is carried out with BFS. An algorithm for finding the connected components of a graph using such a traversal is presented. For any instance of the problem, this algorithm has no greater computational complexity than breadth-first search, and for~most individual problems it has less complexity.

\smallskip

The author thanks V. A. Motovilov, a graduate of the Faculty of Mathematics of Omsk State University, for his help in conducting numerical experiments and valuable comments.\\

\clearpage
\end{document}